# Nature-inspired three-dimensional surface serration topologies enable silent flight by suppressing airfoil-turbulence interaction noise

Zixiao Wei[1,2], Stanley Wang[1,2], Sean Farris[1], Naga Chennuri[1], Ningping Wang[1], Stara Shinsato[1], Kahraman Demir[1], Maya Horii[1], and Grace X. Gu[1*]

[1] Department of Mechanical Engineering, University of California, Berkeley, CA 94720, USA

[2] These authors contributed equally

*Corresponding author: ggu@berkeley.edu

## Abstract

As natural predators, owls fly with astonishing stealth due to the sophisticated serrated surface morphology of their feathers that produces advantageous flow characteristics and favorable boundary layer structures. Traditionally, these serrations are tailored for airfoil edges with simple two-dimensional patterns, limiting their effect on overall noise reduction while negotiating tradeoffs in aerodynamic performance. Here, we formulate new design strategies that can mitigate tradeoffs between noise reduction and aerodynamic performance by merging owl feather and cicada insect wing geometries to create a three-dimensional topology that features silent and efficient flight. Aeroacoustics and aerodynamics experimental results show that the application of our hybrid topology yields a reduction in overall sound pressure levels by up to 9.93% and an increase in propulsive efficiency by over 48.14% compared to benchmark designs. Computational fluid dynamics simulations reveal that the three-dimensional, owl-inspired surface serrations can enhance surface vorticity. The produced coherent vortex structures serve to suppress the source strength of dipole and quadrupole pressure sources at various Reynolds numbers, resulting in a universal noise reduction effect. Our work demonstrates how a bioinspired three-dimensional serration topology refines the turbulence-airfoil interaction mode and improves multiple functionalities of an aerodynamic surface to enable quieter and more fuel-efficient, aerial vehicles.

## Introduction

Smart aeroacoustics design draws great attention due to the rapid popularization of urban usage of aerial vehicles and rising restrictions on noise pollution. Traditionally, designing aeroacoustic propellers, for example, relies on the concept of B-spline[1], where a set of control points generated by nested functions are introduced to formulate an aerodynamic surface. However, this class of formulation is limited by a series of topological constraints, such as the requirement of surface smoothness and continuity. This approach thus creates a ceiling for the further improvement of aeroacoustic performance. Finding an alternative design formulation is challenging especially when considering a three-dimensional design space.

Consequently, researchers have investigated natural predators that are reputed for the stealth of their flight, seeking innovative and effective solutions. Having evolved over billions of years, biological creatures can serve as great design templates for human-made products and engineering systems[2-7]. For instance, creatures such as the owl have developed unique wings and fringe morphologies for quietness[8]. This adaptation has been a foundational element in the design of bioinspired passive noise control devices. Wang et al.[9] provide a review of owl-inspired aeroacoustic devices and identify three distinct features of an owl feather: leading-edge serrations, trailing-edge fringes, and a soft downy (velvet-like) coating. Owl-inspired modifications to aerodynamic surfaces have resulted in a variety of solutions, including variations of leading-edge serrations[10-13], trailing-edge serrations[14], and the application of porous dampening materials[15]. However, in conventional designs such as the leading-edge sawtooth serration, slitted serration[16], and sinusoidal serration, the pursuit of passive noise reduction comes at a penalty of overall aerodynamic performance.

In this work, we formulate new design strategies that can mitigate tradeoffs between noise reduction and aerodynamic performance by merging owl feather and cicada insect wing geometries to create a three-dimensional topology that features silent and efficient flight. Specifically, insect wing geometries such as the cicada have been established as promising templates in the design of aerodynamically advanced devices and propellers[17,18]. Hence, we use the wing shape of the cicada as a complementary planform that addresses the requirements for both aerodynamic and aeroacoustic performance. While preceding works inspired by owl fringe have examined the implementation of simplified, two-dimensional serration modifications, our design features a three-dimensional topology where the serration pattern extends to be a full-surface mutation instead of a local variation along edges. In our design law, two three-dimensional sinusoidal splines are set as the guidelines for lofting, which generates a surface with embedded serration textures shown in **Figure 1(a)**.

To validate our hybrid design strategy combining owl feather and cicada geometries, a hybrid aeroacoustics and aerodynamics measurement system is employed in our experiments. A group of propeller-shaped prototypes consisting of several representative topologies is examined and compared to isolate the effects of owl feathers and cicada features. The topology highlighting our design is the 3D sinusoidal cicada (3D-SC) topology. Our control group has three benchmark designs. A smooth, cicada-shaped topology is set as the first benchmark (B1) to isolate the effect of 3D serrations. In order to signify the effect of the cicada planform, a conventionally shaped, serrated prototype is set to be the second benchmark (B2) while a conventionally shaped, non-serrated prototype is set to be the third benchmark (B3). To this end, the comparison between 3D-SC and B1 prototype reveals the contribution of the 3D surface serrations, while the comparison between B1 and B3 manifests the effect of the cicada planform. The direct comparison between 3D-SC and B3 design defines the overall acoustic and aerodynamic reinforcement associated with the proposed topology. These experimental prototypes are fabricated using Polyjet additive manufacturing processes discussed in the Supporting Information and as shown in **Figure 1(b)**. Moreover, computational fluid dynamics (CFD) simulations are performed to dissect the underlying mechanisms, providing a complementary perspective to experimental findings. The fundamental mechanisms studied in this work can lead to promising applications in a multitude of fields, including urban air mobility, wind power generation, and hydrodynamic vehicles.

## Results and discussion

**Topological design concepts:** In the design process for the 3D-SC propeller, we start by digitalizing the morphology of the owl fringe and cicada wing shapes such that their geometries can be expressed explicitly and integrated smoothly as illustrated in **Figure 1(a)**. The chord ($\mathcal{C}$) layout of the cicada planform with respect to the spanwise position $x$ is first extrapolated through a fifth-order polynomial curve fit of a representative cicada wing profile using leading ($\ell$) and trailing ($t$) edge functions.

$$\mathcal{C}_\ell(x) = -0.007969x^5 + 0.0339x^4 - 0.02455x^3 - 0.167x^2 + 0.4579x + 0.05093 \quad (1)$$

$$\mathcal{C}_t(x) = -0.008017x^5 + 0.04249x^4 + 0.002842x^3 - 0.08049x^2 - 0.04942 \quad (2)$$

3D surface serrations are then added through the superposition of a sinusoidal pattern in the definitive functions of leading and trailing edge ($\mathcal{C}_{\ell/t}$) and shown in **Figure S1(a-c)** in Supporting Information. To characterize the sinusoidal pattern, a dimensionless variable known as aspect ratio ($\Lambda$) is introduced, as shown in Eq (5).

$$\mathcal{C}_{3D-sc} = \mathcal{C}_{\ell/t}(x) + f_s(x) \quad (3)$$

$$\mathcal{C}_s(x) = \mathcal{A} \cdot \sin\left(\frac{2\pi}{\lambda}x\right) \quad (4)$$

$$\Lambda = \frac{\mathcal{A}}{\lambda} \quad (5)$$

where $\mathcal{A}$ and $\lambda$ represent the amplitude and wavelength of the constitutive sinusoidal, $s$, wave function.

Subsequently, the 3D serrated surface is lofted in Computer-Aided Design (CAD) software using the leading and trailing edge curve functions as guidelines. The 2D airfoil used for lofting is NACA 8412, which features a high lift coefficient, lift to drag ratio, and stalls at approximately 25 degrees at the operating Reynolds (Re) numbers, as shown in **Figure S1(d)** in the Supporting Information. Since the maximum chord of a cicada planform is closer to its tip compared to a conventional planform, the cross-span attack angle of the cicada-based prototypes (3D-SC, B1) is fixed to be 15 degrees to prevent flow separation. In order to examine the propellers under a broader range of rotational speeds without inducing structural failure, the rotor diameter is set to be 6 inches, which is smaller than most propellers used for commercial unmanned aerial vehicles (UAV)[19]. It is of note that the limited rotor size leads to a relatively low magnitude of propulsive efficiency (~10%). Additional experiments are conducted to ensure that a 12-inch version of the 3D-SC planform can reach a reasonable efficiency range, as discussed in the Supporting Information.

**Acoustic advancement at various rotational speeds in experiments:** In order to examine the aeroacoustic performance of our designs, we first focus on the composition of rotor noise caused by propellers. In general, rotor noise consists of two primary sources: tonal noise and broadband noise[20]. These two sound sources are driven by different physical mechanisms and exhibit distinctive signatures in the sound spectrum. Specifically, tonal, or harmonic, noise is caused by periodic rotation, presenting as anchored pitches correlated to the rotation frequency. Tonal noise can be further broken down into loading noise and thickness noise. Loading noise is caused by both steady and unsteady aerodynamic loading, while thickness noise is caused by local fluid expansion over the propeller surface. Broadband noise also has different sources of noise generation. The most dominating noise source is blade-wake interaction noise, resulting from the interaction between tip vortices and the propeller blade.[21] To characterize and compare the rotor noise of each topology (3D-SC, B1, B2, and B3), we collect their sound signals under different rotational speeds through an omnidirectional microphone, as listed in **Table 1** (more details in the Supporting Information). The overall sound pressure level (OASPL) curve of each propeller as a function of the rotational speed is obtained through a second-order curve fitting, as shown in **Figure 2(a-b)**. The plot shows that the 3D-SC design possesses an overall lower OASPL compared to the B1, B2, and B3 designs at various rotational speeds. This advancement is more evident at high rotational speeds.

While OASPL directly reflects the associated noise of a propeller, we probe further into the sound spectrum where the sound pressure level is plotted in the frequency domain as shown in **Figure 2(c)**. It is noteworthy that the resultant sound pressure level (SPL) of a rotor is significantly influenced by flow turbulence in near-wall boundary layers. Re number is, hence, introduced to characterize the flow property regarding how easily the potential perturbation can trigger turbulent flow behaviors:

$$Re = \frac{\rho \omega x C}{\mu} cos(\alpha) \tag{6}$$

where $\omega$ denotes the rotor's angular rotational speed in revolutions per minute (RPM), $\alpha$ denotes the effective attack angle towards the cross-sectional airfoil, and $\mu$ represents the dynamic viscosity. The Re number corresponding to each experimental scenario is listed in **Table S1-3** as a reference of the flow's turbulence level (see Supporting Information).

Sound spectrums at 2000 RPM and 5000 RPM are chosen to highlight the acoustic signatures of these propellers under low and high Re numbers, as shown in **Figure 2(c)**. In these two spectrum plots, the 5th-order Savitzky-Golay filter[22] is used to increase the frequency bin width, filter out high-frequency data noise, and highlight the signal features corresponding to tonal and broadband noises.

From experiments, the rotor sound profile at 2000 RPM is dominated by acoustic tones with frequencies lower than 14 kHz. This confirms that harmonic loading is the primary source of propeller noise at low rotational speeds. Notably, the SPLs of multiple tones of a 3D-SC propeller are apparently lower than B1, B2, and B3 prototypes, as shown in the 2000 RPM spectrum plot. At high Re numbers, as shown in the 5000 RPM sound spectrum plot, acoustic tones become less prominent while broadband noises come into dominance. At 5000 RPM, the SPL of a 3D-SC propeller is full-spectrum lower than the other three benchmark designs. The comparison between B1 and B3 prototypes suggests that the cicada planform carries out a trivial acoustic improvement. The difference between B2 and B3, compared to the difference between 3D-SC and B1, manifests that 3D surface serrations result in a more pronounced noise reduction impact on the cicada planform compared to the conventional planform, particularly at higher rotational speeds.

As a result of these differences, the OASPL of the 3D-SC propeller is measured to be 5.14 dB (6.19%) lower than the B1 prototype at 2000 RPM and 9.95 dB (9.57%) lower at 5000 RPM, representing the amount of noise reduction contributed by 3D surface serrations. Regarding the noise reduction effect of the cicada planform, the OASPL of the B1 propeller is 1.62 dB (3.8%) lower than the B3 at 2000 RPM and 0.41 dB (0.39%) lower at 5000 RPM. The synergy of 3D surface serrations and cicada planform leads to a maximum noise reduction of 10.36 dB (9.93%) by comparing the OASPL of the 3D-SC and B3 prototypes at 5000 RPM. At 2000 RPM, the overall OASPL reduction associated with the 3D-SC topology is 6.76 dB (7.98%), which is comparatively lower.

In summary, both cicada planform and 3D serrations contribute to reducing acoustic noise. The contribution from the cicada planform slightly decreases with the increment of rotational speed. In contrast, 3D surface serrations correlate to all-frequency (0-20kHz) attenuation, with the attenuation strength growing with rotational speed. It is noteworthy that the 3D surface serration carries out a more consequential noise reduction effect than the cicada planform.

**Acoustic comparison with leading-edge serrations:** At the early stage of our study, the acoustic performance of the leading-edge serration is compared with the 3D surface serration. In the experiment, two identical, conventionally shaped propellers are respectively reinforced by these two types of serrations. To avoid any potential bias arising from differences in sinusoidal waveforms, we maintained a consistent amplitude and wavelength of 2.5 mm and 10 mm for the 3D serrated prototype. The leading-edge serrated prototypes are designed with different amplitudes and wavelengths (including the 2.5mm×10mm baseline prototype) to form a comprehensive comparison. Through experiments, we discovered that the OASPL of a 3D-serrated propeller is 3.63 dB (4.2%) lower than the leading-edge serrated propeller when their amplitudes and wavelengths are identical (see **Figure S2** in supporting materials for details). Furthermore, the data presented in **Table 2** suggest that the 3D surface serrations outperformed all other leading-edge serrated prototypes in terms of noise reduction.

**Experimental evidence of 3D-SC aerodynamic advantages:** After confirming the aeroacoustic advancement of the 3D-SC design, we proceed to probe any potential trade-offs associated with this improvement. As aforementioned, 2D serrations tend to induce aerodynamic penalties, commonly exhibiting a loss of lift and a gain of drag. Thus, it is an indispensable aspect of our research to inspect the aerodynamic performance of the 3D-SC propeller in addition to its acoustic performance. In the static test of a rotor,[19,23] one can inspect the thrust coefficient, Eq. (7), of a propeller at various rotational speeds to assess the competency of an aerodynamic body in producing thrust. The propulsive efficiency of a rotor is determined by the ratio of the net power output to input of a rotor system, as indicated by Eq. (8). Thus, this efficiency is also referred to as system efficiency, providing a comparatively impartial measurement of the energy converting rate of a propeller. The system efficiency takes into consideration the changes in both mechanical and electrical power, ensuring a comprehensive evaluation (refer to Supporting Information for more details).

$$C_T = \frac{T}{\rho\, n^2\, d^4} \qquad (7)$$

$$\eta = \frac{T\, u_0}{P} \tag{8}$$

where $C_T$, $\eta$, $T$, and $P$ denote the thrust coefficient, propulsive efficiency, thrust generation, and power consumption of the rotor, respectively. The variables $\rho$ and $u_0$ represent the density and velocity of the freestream fluid (assuming air is incompressible at low Re numbers), respectively. Lastly, $n$ and $D$ mark the rotational speed and disk diameter of the rotor, respectively.

As shown in **Figure 2(d),** the thrust generation of the propellers is measured at rotational speeds ranging from 0 to 5500 RPM, and the average power of the propeller is quantified at 5-50 g equivalent thrusts. The thrust coefficient curve indicates that the 3D-SC propeller possesses the highest thrust coefficient among all prototypes at low rotational speeds (<1600 RPM). Compared to B2 and B3, 3D-SC and B1 prototypes share a similar curve form, potentially denoting the characteristics of the cicada planform. As shown in **Figure 2(d)**, the conventional planform produces low thrust coefficients at low rotational speeds, which subsequently increase and plateau at approximately 2500 RPM. In contrast, the cicada planform contributes to high thrust coefficients at low rotational speeds, which decrease and plateau at approximately 2000 RPM. In particular, the 3D-SC propeller exhibits a maximum thrust coefficient that is 0.0124 (23.91%) higher than the B1 propeller and 0.0315 (92.3%) higher than the B2 propeller at 1000 RPM. However, at 3450 RPM, the thrust coefficient of the 3D-SC prototype is considerably lower, with a maximum decrease of 0.0036 (6.2%) compared to B1 and 0.0085 (12.91%) compared to B2. An examination of the Re number reveals that both the 3D surface serration and cicada planform are only effective in generating thrust when the Re number is below 1.43e4.

Despite the reduction of thrust at high rotational speeds, the 3D-SC prototype is remarkably advanced in propulsive efficiency, as shown in **Figure 2(e)**. The box plot at 50 g reveals that the 3D-SC propeller attains the highest efficiency (0.1034) compared to benchmark propellers. Specifically, the propulsive efficiency of the 3D-SC prototype is 0.0139 higher than B1, denoting a 15.50% improvement due to the implementation of 3D surface serrations. Additionally, the propulsive efficiency of the B1 propeller is 0.0197 higher than B3, denoting a 28.21% improvement attributed to the cicada planform. The direct comparison between 3D-SC and B3 prototypes indicates an overall efficiency increment of 48.14%. Notably, the power fluctuations associated with the cicada-based propellers (3D-SC, B1) are higher than the ones using conventional planform (B2, B3), as shown in **Table 3** and **Table S4-5**. The noise detected in the thrust and efficiency measurements is documented in the Supporting Information.

In rotor dynamics, energy dissipates through three different sources of drag: pressure, surface friction, and turbulence. The surface materials and areas of all the testing prototypes are set to be equal, which essentially rules out the difference in surface friction drag. Therefore, the hypothesis posits that the efficiency enhancement stems from the unique chord arrangement of the cicada planform, which introduces a reduced pressure drag. By shifting the maximum chord towards the outer span, the cicadas' wing shape can potentially boost the thrust coefficient, lower the required angle of attack, and thus, mitigate the flow separation that contributes to turbulent drag.

To this end, our empirical results suggest a superiority in both aerodynamic and aeroacoustic performance when combining serration and planform optimization in the sinusoidal cicada propeller design (3D-SC). However, to understand the underlying mechanism behind this advantage, we use computational fluid dynamics (CFD) to investigate the effects of fluid characteristics related to the 3D-SC performance in the next section. Notably, we primarily focus on explaining the aeroacoustics phenomenon in this study. The investigation of the mechanism behind aerodynamic advancement will be covered in future research.

**Investigation of fluid mechanics using computation:** CFD simulations serve to reconstruct the process of rotation based on the given physical constraints, where the amplitude, frequency, and distribution of fluctuating pressure sources dictate the resultant SPL profile. The pressure potential equation of three-dimensional sound wave functions[24] is presented below:

$$\frac{1}{(a^2)_\infty} \frac{\partial^2 p'}{\partial t^2} - \Delta p' = \left(1 - \frac{\sigma_\infty p_\infty}{\rho_\infty T_\infty}\right) \frac{\partial \dot{m}'}{\partial t} + \frac{\sigma_\infty}{T_\infty} \frac{\partial \dot{\vartheta}}{\partial t} - \nabla \cdot \left(\rho^0 \frac{\partial \dot{v}}{\partial t} + \nabla p'\right) \qquad (9)$$

In this equation, all symbols with an ∞ subscript represent properties of the freestream flow, which are considered constants in our simulation. Concerning the mechanics of an incompressible $\left(\frac{\partial \dot{m}}{\partial t} = 0\right)$ fluid in a quasi-adiabatic procedure $\left(\frac{\partial \dot{\vartheta}}{\partial t} = 0\right)$, the strength of pressure fluctuation $\left(\frac{1}{(a^2)_\infty} \frac{\partial^2 p'}{\partial t^2}\right)$ is inversely related to the changing rate of the velocity field $\left(\nabla \cdot \frac{\partial \dot{v}}{\partial t}\right)$ in Cartesian space. In this sense, the local velocity profile changes in a periodic pattern at each rotation cycle, leading to high-intensity sound power allocated to fixed frequencies. These sound signals present as spikes (i.e., acoustic tones) in the sound spectrum. On the other hand, eddies emerge in the turbulent layers located further from the wall, generating sound signals that are neither power intense nor attached to a specific frequency. These sound signals spreading across the entire spectrum manifest as broadband noise. To characterize the acoustic trait of a 3D-SC design, we conduct CFD simulations at both low (2000) RPM and high (5000) RPM to examine the propeller under laminar and turbulent conditions.

Based on streamline plots, shown in **Figure 3(a)**, the serration-reinforced surface exhibits two distinctive flow features compared to a smooth surface. The first being a spanwise, centrifugal airstream, and the second being vortices clustering near the tip and the trailing edges of the blade**.** Compared to turbulent eddies that appears small in scale and exists short in time, coherent vortex structures (CVS) preserve angular momentum of the flow and hence contribute to a lower dissipation rate. This characteristic can effectively reduce the intensity of flow-wall interaction and impact the resultant sound profile. Thus, the identification of CVS is essential to understanding the underlying mechanism. In our study, we present both the iso-helicity and swirling strength contours to assist the visualization of CVS. As an invariant of Euler's equation, helicity[25] is defined as the weighted integral of the vorticity field, which is given by:

$$H = \int_V u \cdot (\nabla \times u) \, dV \qquad (10)$$

where $\nabla \times u$ stands for the curl of the velocity field, while the volume integration quantifies the total amount of rotation within the volume that encloses vortex lines. From the standpoint of topological fluid dynamics, helicity also reflects the linkage[26] and knottedness[27] of vortex lines[28]. Swirling strength is quantified by the swirling number[29], with its expression in polar coordinates defined as:

$$S = \frac{\int r\omega\vec{v} \cdot d\vec{A}}{\bar{R} \int U\vec{v} \cdot d\vec{A}} \tag{11}$$

where $r$ stands for the radial coordinate, $\omega$ stands for the swirl velocity, $\vec{v}$ stands for the radial velocity, and U represents the axial velocity. As the equation indicates, the swirling number is defined as the ratio between the angular and axial momentum of the axial flux. In particular, swirling strength marks the presence of influential vortex structures and their rotational strength.

According to **Figure 3(b)**, the 3D-SC propeller contributes to full-span helicity at 2000 RPM. In contrast, the B1 propeller is vortex-sparse, with vortices distinguished near the tip of the propeller. The swirling strength contour shown in **Figure 3(c)** reveals an analogous vortex layout where the 3DSC prototype produces comparatively more continuous vortex structures. Judging by the swirling strength magnitude, 3D serrations also result in stronger flow rotations compared to a smooth surface. Notably, surface smoothness is conducive to preserving the laminarity of the flow due to the lower level of perturbation. The interaction between laminar flow and the array wall induces steady pressure variation at each charge of loading, which causes the growth of dipole pressure sources (i.e., tonal noises). In this sense, smoothness is not desired for tonal noise reduction at low Re numbers. On the contrary, the uneven surface of a 3D-SC propeller introduces additional contact roughness. The valley-like, serrated surface configuration delivers a localized pressure difference (shown by **Figure S3** in Supporting Information) which is hypothesized to drive the formation of CVS, as **Figure 3(b-c)** shows. CVS presents as continuous and large topologies in a frozen fluid domain, contributing to longer persistence in time[30]. This characteristic maintains the inertia of CVS along the motion direction and prevents the flow from breaking down into small-scale eddies. This mechanism aids in suppressing the turbulent load (i.e., pressure strength corresponding to broadband noises), as indicated by the cross-comparison between the helicity and quadrupole pressure contour. On the other hand, the rotation associated with longitudinal vortex lines periodically lifts the flow away from the boundary wall[31], reducing the intensity of harmonic aerodynamic load in flow-wall interaction (i.e., dipole pressure strength corresponding to tonal noises). Therefore, the production of CVS contributes to suppressing both tonal and broadband noises, as illustrated in **Figure 3(d-e).** This explains why the average SPL value and the SPLs of multiple acoustic tones of the 3D-SC propeller are measured to be lower than any other benchmark prototypes at 2000 RPM.

At high Re number (5000 RPM), the inertial load upon the fluid overshadows the viscous load, turning any small perturbations into turbulence (see **Figure S4** in the Supporting Information). This transition is evident in the sound spectrum, as tonal noises gradually become less pronounced amidst broadband noises, accompanied by a surge in the average SPL. By comparing the helicity and swirling contours of the 3D-SC propellers at 2000 and 5000 RPM in **Figure 3(b-c)**, we notice that structured vortices cannot withstand excessive disruption and decompose as the Re number increases. Despite the overall abatement, the contour indicates that the 3D-SC propeller still produces a broader helicity region than the B1 prototype at 5000 RPM, with extra vortices generated around the tip and the trailing edge of the propeller. By comparing the sound pressure contours of the 3D-SC propeller at 2000 and 5000 RPMs in **Figure 3(d-e)**, it is noteworthy that the pressure levels of both dipole and quadrupole sources increase significantly. Judging by the pressure magnitude, the quadrupole (broadband) pressure sources come into dominance at 5000 RPM. On this premise, the 3D-SC propeller correlates to a significant suppression of quadrupole pressure near the blade's trailing edge, as shown in **Figure 3(e)**, which coincides with the vortex-dense region shown in the helicity contour. Based on this observation, it is believed that the raised vorticity due to 3D surface serrations is responsible for the reduction of broadband noise, which agrees with our observation in acoustic data.

To this end, our computational results show that implementing 3D surface serrations can passively reduce noises at various rotational speeds because of the continuous and localized pressure difference created by the valley-like surface morphology. The resultant pressure distribution encourages the formation of CVS across the boundary layer that tends to weaken the acoustic tones associated with laminar flow at low Re numbers. Additionally, it helps maintain a low broadband noise by preserving the inertia of the fluid along its motion direction, thus withholding the decomposition of large CVS into small turbulent eddies at high Re numbers.

**Parametric study of constitutive sinusoidal wave functions:** The noise attenuation of a 3D-SC propeller originates from its 3D surface serration. Naturally, formulation of the serration geometry determines the resultant performance metrics. Unlike two-dimensional serrations, design parameters of three-dimensional serrations have a more substantial impact on the resulting topology, as the sinusoidal patterns extend across the entire surface. Moreover, the size of surface textures can potentially influence the scale of generated CVS, affecting the consequent sound attenuation effect. Therefore, we pursue a parametric optimization to produce a greater improvement of quietness for a 3D-SC propeller. Proceeding with this systematic optimization, sixteen combinations of amplitude and wavelength are investigated in our study. As shown in **Figure 4(a)**, the testing matrix consists of different sinusoidal patterns with their amplitudes varying from 0.01 to 0.04 inches and wavelengths ranging from 0.1 to 0.4 inches, representing the transition from smooth to densely serrated surface. The experimental data of thrust and OASPL for these propellers are collected from 2000 to 6000 RPM (shown in **Table S6-7** in Supporting Information). Moreover, a high-order surface interpolation is employed to construct the contour plot of OASPL and thrust with respect to the design parameters, as shown in **Figure 4(b)**.

In these contour plots, we observe that there are equipotential lines lying diagonally in the plot, where the changing rate of amplitude with respect to wavelength is constant. Moreover, the OASPL and thrust measurement of the data points along an equipotential line are numerically close. Particularly, we find that the properties of a 3D-SC propeller are determined by the slope of an equipotential line and its intersection point with the $y$-axis. Namely, the performance metrics of a 3D-SC propeller are dictated by a dimensionless parameter (i.e., the aspect ratio of the sinusoidal pattern) and a dimensional parameter (i.e., size of the sinusoidal feature). At 2000 RPM, the OASPL and thrust of the 3D-SC propeller both reach their minima at the highest aspect ratio (wavelength of 0.1 inch and amplitude of 0.04 inches). As demonstrated in Figure 4A, this prototype is constructed with the highest serration density. At low RPM, the laminar flow is replaced by vortices, which leads to the destruction of acoustic tones and the accompanying fall of lift due to the loss of dynamic pressure into driving flow circulation. At 5000 RPM, the thrust of a 3D-SC propeller remains minimum at the highest aspect ratio. The OASPL, however, reaches its maximum. This opposite trend indicates that a dense three-dimensional structure can augment the noise at high Re numbers. When two adjacent serrations are located sufficiently close, the localized pressure gradient gets repressed, and the serration turns into another source of perturbation that increases the broadband noise. It is also noteworthy that the OASPL at 5000 RPM remains invariant with the design parameters when the amplitude is lower than 0.25 inches. As the amplitude exceeds above 0.25 inches, the OASPL consistently increases with the aspect ratio. It is hypothesized that the serration size should be sufficiently large at high Re numbers to invoke the fluid mechanics for noise reduction.

**Conclusions**: In summary, we discover a 3D-SC surface topology that reduces the flow-blade interaction noise and improves the aerodynamic performance of a propeller design at various rotational speeds. The novel 3D sinusoidal serration leads to a full-spectrum reduction of SPL and a 3.63 dB (4.24%) reduction of OASPL at 4000 RPM compared to leading-edge sinusoidal serrations, denoting the benefit of extending 2D plane serrations to 3D surface textures. Furthermore, it is revealed by experiments that the synergy between two biological creatures (cicadas and owls) leads to aerodynamic and acoustic amelioration that the morphology of a single creature cannot provide in isolation. Particularly, the 3D-SC planform carries out a maximum OASPL reduction of 6.76 dB (7.98%) at 2000 RPM and 10.25 dB (9.93%) at 5000 RPM. Regarding the aerodynamic metrics, the 3D-SC prototype maintains an advancement in thrust coefficient at low rotational speeds (<1600 RPM). At higher rotational speeds, the cicada planform results in a maximum thrust coefficient loss of 12.91%. Nonetheless, propulsive efficiency is remarkably improved. Specifically, we observe a 15.5% efficiency increment associated with the 3D surface serration and a 28.21% increment for the cicada planform, leading to an overall efficiency improvement of 48.14% attributed to the 3D-SC topology. In the exploration of underlying mechanisms, CFD simulations reveal that the 3D serrations contribute to producing CVS over the surface of a propeller. The vortex lifting effect associated with CVS suppresses the harmonic aerodynamic loading at low Re numbers, undermining the dipole pressure strength corresponding to tonal noises. At high rotational speeds, large-scale vortex structures defer the transition into turbulence and thereby reduce the strength of quadrupole pressure sources corresponding to broadband noises. Based on this mechanism, it is hypothesized that the geometric parameters of 3D serrations can affect the scale of CVS being produced and consequently affects the attenuation of sound signals. Through experiments, parametric study manifests that the 3D-SC propeller is sensitive to the amplitude and wavelength of the constituent sinusoidal pattern. Remarkably, when the propeller is densely serrated (high amplitude to wavelength ratio), the resultant topology leads to the lowest noise level. In this sense, the sound attenuation ability of the 3D serration topology can be formulated by changing the design parameters. To this end, the 3D-SC topology could turn passive noise control into active attenuation as a future study.

**Experimental section**

Experimental and computational details can be found in the Supporting Information.

**Acknowledgements**

This research was supported by CITRIS and the Banatao Institute, Air Force Office of Scientific Research (Fund number: FA9550-22-1-0420), and National Science Foundation XSEDE Supercomputing Resources (Fund Number: ACI-1548562).

**Tables**

**Table 1.** Statistics of propulsive efficiency measurement

| Rotational speed (RPM) | OASPL (dB) | | | |
|---|---|---|---|---|
| | **3D-SC** | **B1** | **B2** | **B3** |
| 1000 | 77.99 | 83.13 | 79.25 | 84.75 |
| 2000 | 79.53 | 84.14 | 81.40 | 87.01 |
| 3000 | 83.08 | 89.99 | 84.11 | 91.15 |
| 4000 | 85.37 | 94.67 | 88.49 | 98.70 |
| 5000 | 94.02 | 103.97 | 96.52 | 104.38 |

**Table 2:** OASPL comparisons between leading-edge serrations and 3D surface serrations

| Serration Type | λ(mm) | A(mm) | OASPL (dB) |
|---|---|---|---|
| Leading edge | 15 | 2.5 | 86.64 |
| Leading edge | 15 | 2.0 | 86.74 |
| Leading edge | 10 | 6.5 | 86.99 |
| Leading edge | 10 | 2.5 | 85.44 |
| Leading edge | 5 | 8.0 | 83.00 |
| Leading edge | 5 | 2.5 | 83.36 |
| 3D surface | 10 | 2.5 | 81.81 |

**Table 3:** Statistics of propulsive efficiency measurement at 50g equivalent thrust

| Properties | Prototype designs | | | |
|---|---|---|---|---|
| | **3D-SC** | **B1** | **B2** | **B3** |
| $\bar{\eta}$ | 0.1034 | 0.0895 | 0.0728 | 0.0698 |
| $\sigma_\eta$ | 0.0413 | 0.0193 | 0.0038 | 0.0033 |
| noise level | 39.9% | 21.6% | 5.3% | 4.7% |

**Figures and captions**

[a]

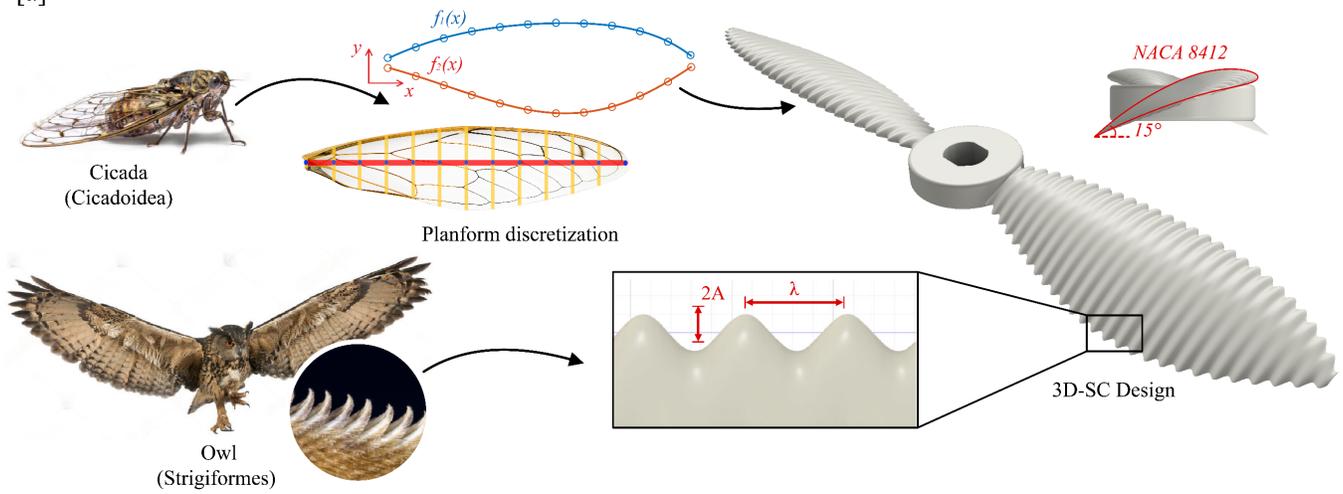

[b]

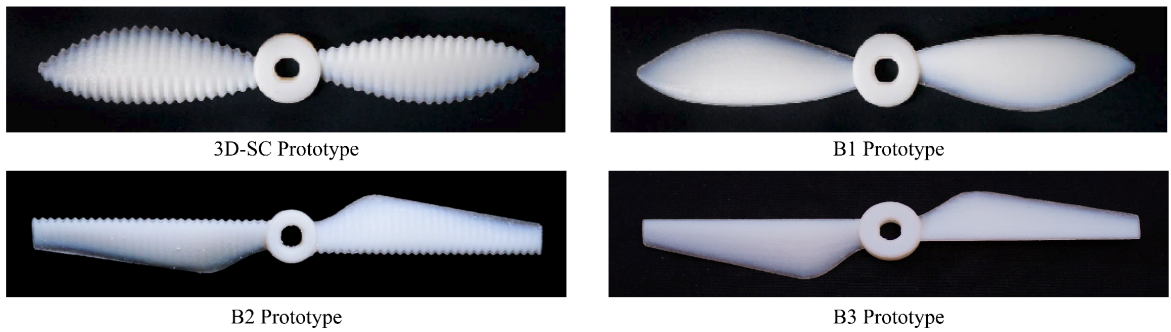

**Figure 1.** Illustration of the design concepts for a 3D-SC propeller. (a) 3D-SC topology inspired by owl fringe and cicada wing shape. (b) Fabricated 3D-SC, B1, B2, and B3 prototypes using polyjet additive manufacturing.

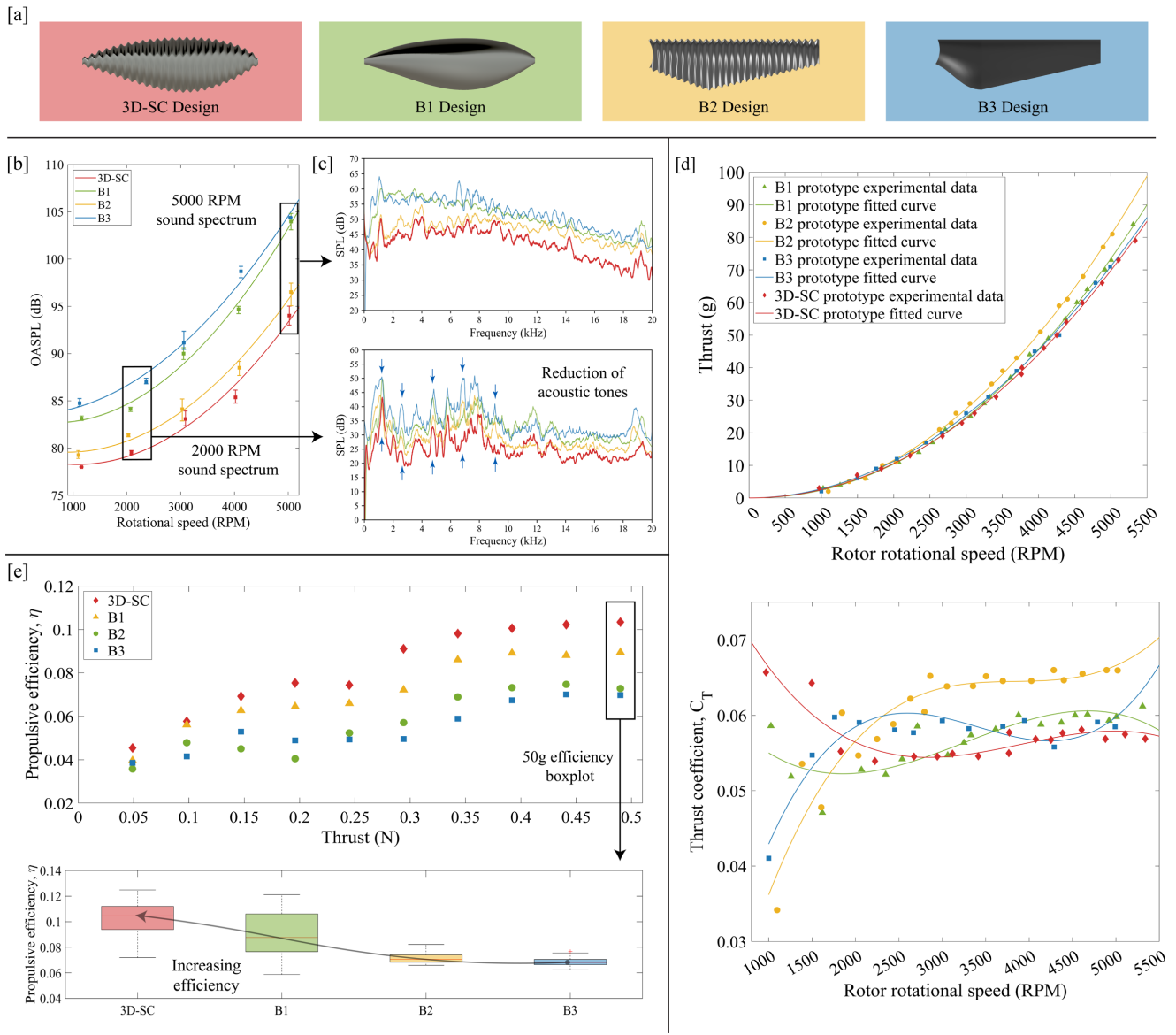

**Figure 2.** Aerodynamic and aeroacoustic experimental results. (a) Schematic CAD models of 3D-SC, B1, B2, and B3 prototypes with corresponding color representations. (b) OASPL measurements from 1000 to 5000 RPM. (c) SPL measurements at 2000 and 5000 RPM. (d) Raw thrust data, fitted thrust curves (top), and thrust coefficient against rotational speeds (bottom). (e) Propulsive efficiency of all prototypes at various thrusts (top) and efficiency box plot at 50 g equivalent thrust (bottom).

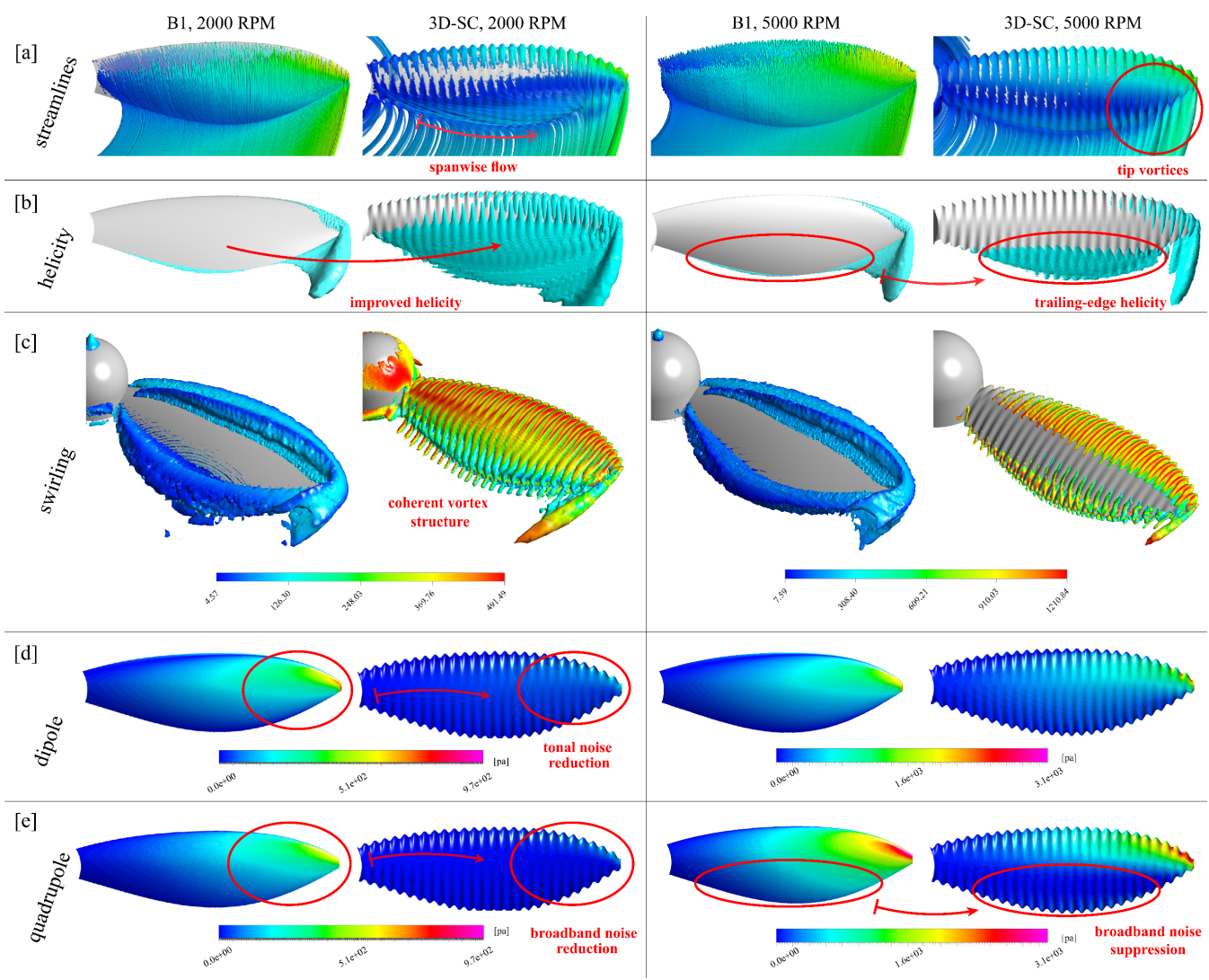

**Figure 3.** CFD simulation results of 3D-SC and B1 topologies corresponding to high (5000 RPM) and low (2000 RPM) turbulence conditions. (a) Streamline plot, (b) helicity contour, (c) swirling strength contour, (d) dipole pressure contour, and (e) quadrupole pressure contour.

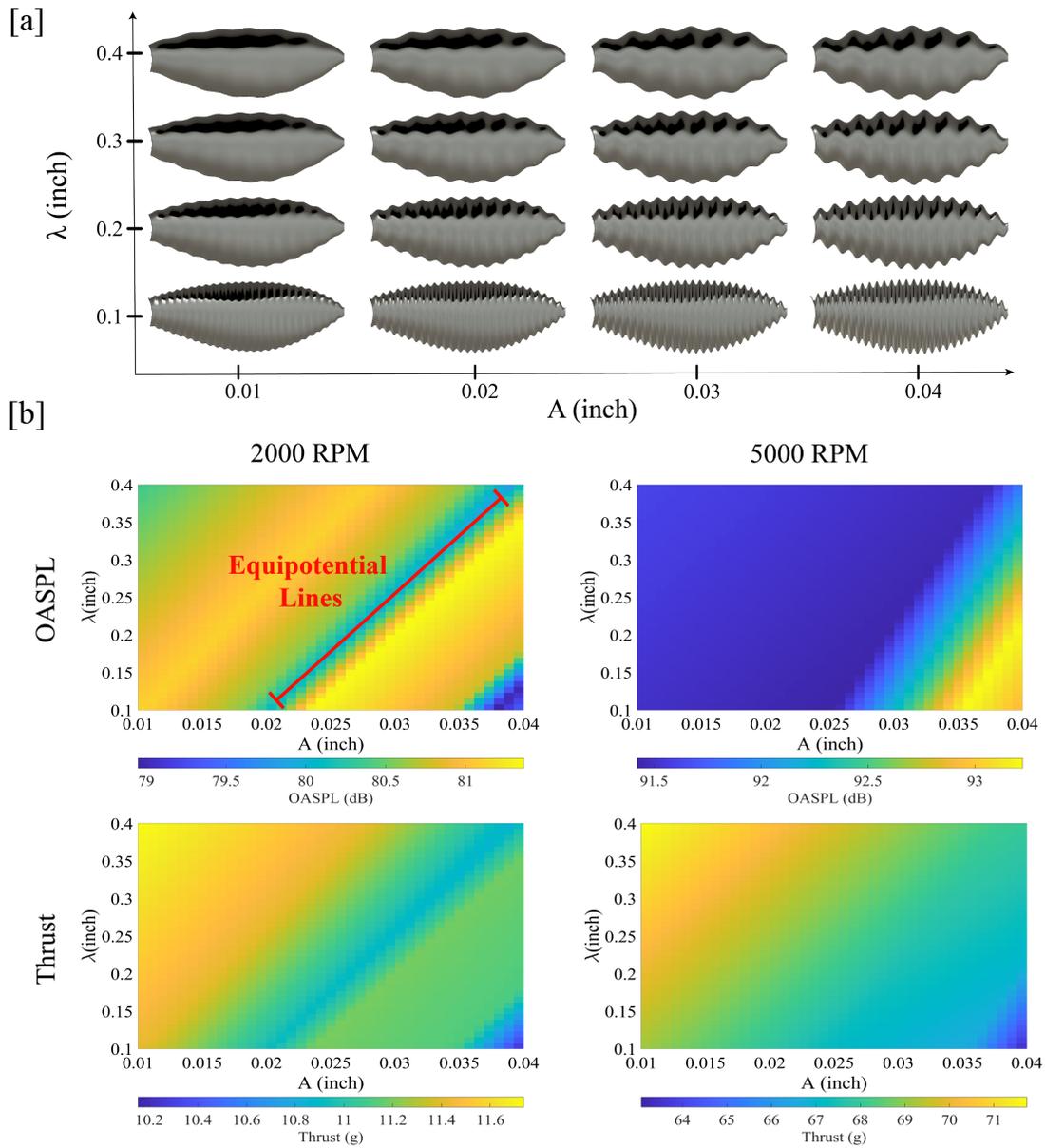

**Figure 4.** Parametric optimization of the constituent sinusoidal pattern of a 3D-SC design. (a) Design space of varying 3D-SC prototypes. (b) OASPL and thrust data of varying 3D-SC prototypes at different rotational speeds.

# Supporting information

# Nature-inspired three-dimensional surface serration topologies enable silent flight by suppressing airfoil-turbulence interaction noise

Zixiao Wei[1,2], Stanley Wang[1,2], Sean Farris[1], Naga Chennuri[1], Ningping Wang[1], Stara Shinsato[1], Kahraman Demir[1], Maya Horii[1], and Grace X. Gu[1*]

[1] Department of Mechanical Engineering, University of California, Berkeley, CA 94720, USA

[2] These authors contributed equally

*Corresponding author: ggu@berkeley.edu

## Methods

**Experimental setup:** A traditional setup for powering and controlling a drone motor and propeller requires the following components: a battery, a flight controller, an electronic speed controller (ESC), the motor, and the propeller. For our experimental test, we utilize an HRB 6000 mAh 3S LiPo Battery, a Racerstar Motor Thrust Stand V3, a 30 Amps RC Brushless Motor Electric Speed Controller, and Readytosky 2212 920KV Brushless Motors. The Racerstar thrust stand includes a built-in high-precision thrust sensor with a maximum magnitude of 5 kg to measure the thrust generation and serves the same role as the flight controller, allowing us to adjust the speed of the motor. In addition to the thrust stand, we leverage a NEIKO 20713A Digital Tachometer to measure the rotational speed in RPMs, a miniDSP UMIK-1 omnidirectional USB microphone to measure the acoustic data, and an RGBZONE 200 Amps RC Watt Meter to measure the power draw into the ESC.

The testing setup consists of a projector stand for the thrust stand setup, a light stand to mount the tachometer, and a microphone stand to hold the microphone (as shown in **Figure S5**). To collect the acoustic data of the rotor, the microphone is placed directly in front of the motor 1 meter away. Both the thrust stand and tachometer have built-in monitors that show the result directly. We use the REW Room EQ Wizard (REW) as an acoustic measurement software with automatic SPL calibration and discovery features to record sound data. The SPL data is obtained at a sampling rate of 48 kHz with no timing difference in the sweep. In order to minimize measurement discrepancy, three sound measurements are taken for each rotational speed and the average of these measurements is taken as the reference value.

The overall sound pressure level represents the total energy that is contained across all resolved frequencies and can be used to provide a singular measurement for the intensity of the noise. Two methods can be used to calculate the overall sound pressure level, taking the integral over all resolved frequencies, as seen in Eq. (1) or by taking the root sum square (RSS) pressure and converting it back to decibels. After confirming that both methods resulted in the same result, the latter was chosen as it was more computationally straightforward.

$$OASPL = 10 \cdot log_{10} \int_0^{f_u} 10^{0.1 SPL(f)} df \qquad (1)$$

Concerning measurement uncertainty in experiments, thrust measurements display minimal error and negligible hysteresis (on the order of ±0.1 N). The propulsive efficiency is attained by measuring the motor input power within a 30-second time window. Each input power measurement consists of 60 samples. Constrained by the unsteady current-loop charge, power readings show a greater standard deviation (approximately 5-40% compared to the statistical average) in the real-time measurement (see **Table S4-5**), but statistical averaging allows trends to be extracted explicitly.

**Fabrication:** In our experiments, the testing prototypes require high stiffness performance to survive the excessive aerodynamic loading. A good surface finish is also needed to eliminate the unwanted interference from surface friction drag. Due to the complexity of the topologies, the propellers parts are 3D printed using the Polyjet additive manufacturing process by Stratasys, which has a high-quality surface finish (≈32-micron layer thickness) and the selection of the material offers a range of material properties. Moreover, this technique eliminates the need for post-curing the parts, and print times are reduced to 30 minutes per propeller which helps to print multiple parts at once. Digital ABS materials provide a stiffer propeller and are thus chosen as the material choice for the propellers. The combination chosen for post-processing our propellers are first wet sanding, followed by epoxy coating, and lastly a layer of spray paint. Although mineral oil and wax help in smoothing the part's surface, these techniques do not add noticeable value to material strength. One challenge encountered when applying epoxy and spray paint is that these coatings can concentrate at the edges of the print and may modify the geometry of the 3D sinusoidal serrations. To overcome this, we keep the layer thickness small and dry the part in different orientations to prevent clumping. Using this combination of post-processing techniques, the propeller has smoother surface finish, minimal gaps within the part, and stronger properties for better noise and aerodynamic performance.

**Reynolds number reference table for experiments:** Reynolds number is a direct indicator for the fluid's turbulence level, which varies with the incoming flow speed and hence varies with the chord position. The average, maximum, and standard deviation of the Reynolds numbers at various chord positions under different rotational speeds are computed and listed in **Table S1**. The Reynolds numbers computed at different chord positions at 2000 and 5000 RPMs are listed in **Table S2-3**.

Table S1: Reynolds number at various RPMs

| $\omega$ | Avg Re | Max Re | Std Re |
|---|---|---|---|
| 2000 | 7.499E+03 | 1.432E+04 | 5.963E+03 |
| 3000 | 1.125E+04 | 2.148E+04 | 8.944E+03 |
| 4000 | 1.500E+04 | 2.864E+04 | 1.193E+04 |
| 5000 | 1.875E+04 | 3.580E+04 | 1.491E+04 |
| 6000 | 2.250E+04 | 4.296E+04 | 1.789E+04 |

Table S2: Reynolds number at various chord positions at 2000 RPM

| Chord (m) | Tangential Speed (m/s) | Chord (m) | Reynolds Number |
|---|---|---|---|
| 0.00 | 0.00 | 0.0023 | 0.00E+00 |
| 0.01 | 2.09 | 0.0092 | 1.37E+03 |
| 0.02 | 4.19 | 0.0150 | 4.45E+03 |
| 0.03 | 6.28 | 0.0292 | 1.30E+04 |
| 0.04 | 8.38 | 0.0208 | 1.23E+04 |
| 0.05 | 10.47 | 0.0193 | 1.43E+04 |
| 0.06 | 12.57 | 0.0135 | 1.20E+04 |
| 0.07 | 14.66 | 0.0024 | 2.49E+03 |

Table S3: Reynolds number at various chord positions at 5000 RPM

| Chord (m) | Tangential Speed (m/s) | Chord (m) | Reynolds Number |
|---|---|---|---|
| 0.00 | 0.00 | 0.0023 | 0.00E+00 |
| 0.01 | 5.24 | 0.0092 | 3.41E+03 |
| 0.02 | 10.47 | 0.0150 | 1.11E+04 |
| 0.03 | 15.71 | 0.0292 | 3.25E+04 |
| 0.04 | 20.94 | 0.0208 | 3.09E+04 |

| 0.05 | 26.18 | 0.0193 | 3.58E+04 |
| 0.06 | 31.42 | 0.0135 | 3.00E+04 |
| 0.07 | 36.65 | 0.0024 | 6.23E+03 |

**System efficiency:** The concept of efficiency adopted in our work is referred to as system efficiency, defined as the ratio between the system's net power output and input. A rotor's system efficiency is the product of motor and propeller efficiency[1]. Relevant definitions are listed as follows:

$$\text{Mechanical Power} = \text{Torque} * \text{Rotational Speed} \tag{2}$$

$$\text{Electrical Power} = \text{Voltage} * \text{Current} \tag{3}$$

$$\text{Motor Efficiency} = \frac{\text{Mechanical Power}}{\text{Electrical Power}} \tag{4}$$

$$\text{Propeller Efficiency} = \frac{\text{Thrust} * \text{Freestream Flow Speed}}{\text{Mechanical Power}} \tag{5}$$

$$\text{System Efficiency} = \text{Motor Efficiency} * \text{Propeller Efficiency} \tag{6}$$

Notably, the electrical power varies with the operational torque and drag. Therefore, propeller and motor efficiencies are both influenced by the propeller's aerodynamic properties. In this sense, system efficiency provides a comprehensive measurement of a rotor's aerodynamic performance.

**Table S4**: Average electrical power measurements at various thrusts

| Thrust | Power Input Measurement Mean (W) | | | |
|---|---|---|---|---|
| | 3DSC | B1 | B2 | B3 |
| 5 | 0.92 | 0.95 | 0.90 | 1.01 |
| 10 | 2.15 | 2.18 | 3.10 | 2.98 |
| 15 | 3.46 | 3.58 | 6.34 | 4.20 |
| 20 | 4.70 | 5.39 | 10.81 | 6.73 |
| 25 | 6.78 | 7.49 | 11.64 | 9.30 |
| 30 | 7.20 | 8.92 | 13.82 | 12.39 |
| 35 | 8.50 | 9.53 | 14.46 | 13.25 |
| 40 | 10.30 | 11.20 | 17.16 | 14.31 |
| 45 | 12.07 | 13.76 | 19.70 | 16.26 |
| 50 | 13.86 | 15.40 | 22.98 | 19.51 |

**Table S5**: Standard deviation of electrical power measurements at various thrusts

| Thrust (g) | Power Input Measurement Standard Deviation (W) | | | |
|---|---|---|---|---|
| | 3DSC | B1 | B2 | B3 |
| 5 | 0.22 | 0.14 | 0.08 | 0.19 |
| 10 | 0.23 | 0.23 | 0.38 | 0.29 |
| 15 | 0.39 | 0.48 | 1.07 | 0.53 |
| 20 | 0.64 | 0.56 | 1.08 | 0.56 |
| 25 | 0.90 | 0.89 | 1.41 | 1.39 |
| 30 | 0.82 | 0.89 | 1.23 | 1.14 |
| 35 | 0.98 | 1.36 | 1.31 | 1.20 |
| 40 | 0.57 | 1.75 | 1.58 | 1.29 |
| 45 | 1.59 | 0.77 | 1.05 | 0.99 |
| 50 | 5.54 | 3.33 | 1.23 | 0.94 |

As discussed in the manuscript, the testing prototypes have a 6-inch rotor diameter, which is limited by the 3D printed material strength. The system efficiency is measured to be a low value because of the small rotor size. Though the trend is clearly revealed by the efficiency data shown in **Figure 2(e)**, it is necessary to examine the efficiency on a larger scale rotor to see the trend at larger propeller sizes. In the supplemental tests, 3D-SC and B1 prototypes have been transformed into 12-inch planforms, as shown in **Figure S6(a)**. Notably, the maximum rotational speed of these larger propellers is 2000 RPM (~70g equivalent thrust) to avoid catastrophic structural failure. The efficiency of large prototypes is collected and presented in **Figure S6(b)**. Judging by the OASPL comparison, the larger 3D-SC planform remains quieter than the larger B1 prototype. Furthermore, the efficiency of the 3D-SC prototype grows from 23.6% to 42.2% with the increment of RPM from 1000-2000 RPM. The study conducted by Brandt et al. (2011) tested 79 propellers for small UAV usage with the rotor diameter ranging from 9 to 11 inches[2]. The efficiency range of these propellers was found to be 28% to 65%, measured under various rotational speeds from 3000 to 6000 RPM. In this sense, the efficiency magnitude of our 12-inch 3D-SC propeller at 2000 RPM is considered reasonable.

**Computational simulations**: CFD analysis is used as a tool to theoretically explain the effectiveness of our design. Therefore, a high-fidelity level is important to ensure the reference value of the CFD analysis. In general, our CFD model consists of mesh setup, ANSYS CFX setup, and ANSYS CFX-post process. CFX is an inbuilt module of Ansys Workbench that allows the user to simulate the flow field model of interest. To create a mesh for the propeller, first, the geometry of the part is modified. Cylindrical rotation and stationary fluid regions are added to the propeller geometry. In general, the rotational region represents the section of the fluid domain where the propeller rotates, and the stationary region is part of the fluid domain where the air particles are initialized to be static to simulate the neighboring fluid domain of the rotor. The rotational region encloses the propeller and has a cushion radius of 10 mm with a cushion height of 10 mm (front and back). In total, 1,928,067 nodes and 1,073,0226 tetrahedron elements are used to construct the rotational domain and most of the meshes are allocated to the inflation layers adjacent to the propeller surface. The static region enclosing the rotational region has a cushion radius of 450 mm and a cushion distance of 500 mm to the inlet and 600 mm to the outlet. In total, 388,852 nodes and 2,256,442 tetrahedron elements are used to construct the static region (see more details in **Figure S7**). The size of the domain is chosen via iterative simulations until numerical convergence and a steady pressure profile in both the freestream and wake region are achieved. The reference pressures of both regions are set to be 1 atm to keep in accordance with lab conditions. In terms of the boundary conditions, the fluid domain consists of a relative total pressure inlet of 0 atm, a relative total pressure outlet of 0 atm, general interfaces connecting the rotational and stationary fluid regions, non-slip propeller surface wall, and a cylindrical slip wall (i.e., zero shear stress) enclosing the static fluid region. The frame change of these connection interfaces is set to be a frozen rotor with a pitch angle of 360 degrees.

In addition to the domain and mesh setup, a high-fidelity turbulence model is also indispensable to the validity of CFD analysis. In CFD simulations, we use both the Menter Shear Stress Transport[3] (SST) model and the Large Eddy Simulation[4] (LES) model to formulate the flow turbulence. The simulation results from these two models are in good agreement, with an average absolute deviation of 4.76%. In particular, the Menter SST model leverages blending functions to separate the far-field and near-wall flow domains, exploiting Reynolds Averaged Navier Stokes (RANS) equations that describe the flow regime by its mean properties and additional fluctuation terms. By applying time-average to the disturbance term, Menter SST serves as a low-by-pass filter to retain the sound pressures under the cut-off frequency. Owing to this nature, the model relies heavily on the assumption that eddies are homogeneous and isotropic in turbulent sublayers. On the contrary, LES allows more sophisticated filtering operations to preserve the features of large vortices while stochastically approximating fluctuations that are smaller than the designated cut-off scale. Theoretically, the equality of these two turbulence models indicates that the eddies within the flow field are homogenous at various scales.

**OASPL and thrust data for parametric study**: This section shows the raw OASPL and thrust data measured in the parametric study. For amplitude (A) denotation, the number postfix multiplying 0.01 inch yields the actual size. For wavelength ($\lambda$) denotation, the number postfix is weighted by 0.1 inch instead.

Table S6. OASPL data of the prototypes in the parametric study

| Design Parameter | RPM | OASPL (dB) | RPM | OASPL (dBA) | RPM | OASPL (dBA) | RPM | OASPL (dBA) | RPM | OASPL (dBA) |
|---|---|---|---|---|---|---|---|---|---|---|
| A1 $\lambda$1 | 2030 | 81.34 | 2995 | 82.27 | 4007 | 84.80 | 5070 | 88.63 | 5970 | 90.45 |
| A1 $\lambda$2 | 2050 | 78.77 | 2981 | 81.40 | 4020 | 85.05 | 5015 | 88.06 | 6024 | 91.11 |
| A1 $\lambda$3 | 2016 | 79.33 | 3000 | 81.97 | 4031 | 85.51 | 5023 | 88.67 | 6024 | 91.23 |
| A1 $\lambda$4 | 2059 | 80.22 | 3000 | 83.02 | 4000 | 85.40 | 5020 | 89.39 | 6020 | 90.73 |
| A2 $\lambda$1 | 2008 | 79.04 | 3050 | 82.36 | 3970 | 85.80 | 5000 | 89.11 | 6024 | 91.53 |
| A2 $\lambda$2 | 2030 | 79.13 | 3000 | 83.83 | 4000 | 86.09 | 5010 | 88.76 | 6023 | 91.16 |
| A2 $\lambda$3 | 2008 | 79.35 | 3050 | 81.96 | 3970 | 85.67 | 5000 | 89.13 | 6024 | 91.68 |
| A2 $\lambda$4 | 2045 | 80.04 | 3005 | 82.72 | 4014 | 85.72 | 5010 | 89.89 | 6023 | 91.49 |
| A3 $\lambda$1 | 2054 | 80.10 | 3068 | 81.67 | 4048 | 85.33 | 5041 | 89.02 | 6024 | 91.21 |
| A3 $\lambda$2 | 2024 | 80.48 | 3029 | 84.45 | 3985 | 86.08 | 5017 | 89.79 | 6024 | 93.20 |
| A3 $\lambda$3 | 2023 | 80.66 | 3001 | 82.74 | 4040 | 87.09 | 5040 | 90.46 | 6024 | 92.58 |
| A3 $\lambda$4 | 2023 | 80.45 | 3000 | 84.28 | 4037 | 86.17 | 5033 | 89.55 | 6024 | 91.97 |
| A4 $\lambda$1 | 2023 | 80.54 | 2970 | 83.91 | 4015 | 85.77 | 5042 | 90.17 | 6024 | 91.24 |
| A4 $\lambda$2 | 2024 | 78.51 | 2970 | 82.63 | 4030 | 86.09 | 5035 | 90.39 | 6024 | 93.48 |
| A4 $\lambda$3 | 2054 | 79.68 | 3029 | 84.30 | 4005 | 86.92 | 5030 | 90.13 | 6024 | 92.48 |
| A4 $\lambda$4 | 2004 | 80.18 | 2957 | 81.81 | 4031 | 86.06 | 5022 | 91.37 | 5960 | 93.11 |

Table S7. Thrust measurements of the prototypes in the parametric study

| Design Parameter | RPM | Thrust(g) | RPM | Thrust(g) | RPM | Thrust(g) | RPM | Thrust (g) | RPM | Thrust (g) |
|---|---|---|---|---|---|---|---|---|---|---|
| A1 $\lambda1$ | 2030 | 11.13 | 2995 | 25.04 | 4007 | 45.41 | 5070 | 73.13 | 5970 | 100.96 |
| A1 $\lambda2$ | 2050 | 11.60 | 2981 | 25.24 | 4020 | 45.18 | 5015 | 72.56 | 6024 | 100.30 |
| A1 $\lambda3$ | 2016 | 11.70 | 3000 | 24.74 | 4031 | 44.99 | 5023 | 73.17 | 6024 | 101.02 |
| A1 $\lambda4$ | 2059 | 11.54 | 3000 | 25.55 | 4000 | 46.13 | 5020 | 73.89 | 6020 | 103.01 |
| A2 $\lambda1$ | 2008 | 12.07 | 3050 | 25.62 | 3970 | 45.55 | 5000 | 73.61 | 6024 | 103.17 |
| A2 $\lambda2$ | 2030 | 10.75 | 3000 | 24.81 | 4000 | 42.03 | 5010 | 73.26 | 6023 | 96.78 |
| A2 $\lambda3$ | 2008 | 11.22 | 3050 | 24.51 | 3970 | 43.57 | 5000 | 73.63 | 6024 | 98.79 |
| A2 $\lambda4$ | 2045 | 11.24 | 3005 | 25.92 | 4014 | 43.92 | 5010 | 74.39 | 6023 | 101.12 |
| A3 $\lambda1$ | 2054 | 11.39 | 3068 | 24.60 | 4048 | 43.89 | 5041 | 73.52 | 6024 | 98.82 |
| A3 $\lambda2$ | 2024 | 11.30 | 3029 | 25.21 | 3985 | 43.88 | 5017 | 74.29 | 6024 | 97.18 |
| A3 $\lambda3$ | 2023 | 10.93 | 3001 | 24.48 | 4040 | 42.36 | 5040 | 74.96 | 6024 | 96.81 |
| A3 $\lambda4$ | 2023 | 11.03 | 3000 | 24.28 | 4037 | 44.00 | 5033 | 74.05 | 6024 | 97.83 |
| A4 $\lambda1$ | 2023 | 10.97 | 2970 | 24.13 | 4015 | 43.69 | 5042 | 74.67 | 6024 | 97.28 |
| A4 $\lambda2$ | 2024 | 10.15 | 2970 | 21.89 | 4030 | 40.00 | 5035 | 74.89 | 6024 | 90.04 |
| A4 $\lambda3$ | 2054 | 10.71 | 3029 | 23.07 | 4005 | 42.47 | 5030 | 74.63 | 6024 | 94.89 |
| A4 $\lambda4$ | 2004 | 11.25 | 2957 | 24.46 | 4031 | 42.77 | 5022 | 75.87 | 5960 | 96.76 |

**Supporting figures**

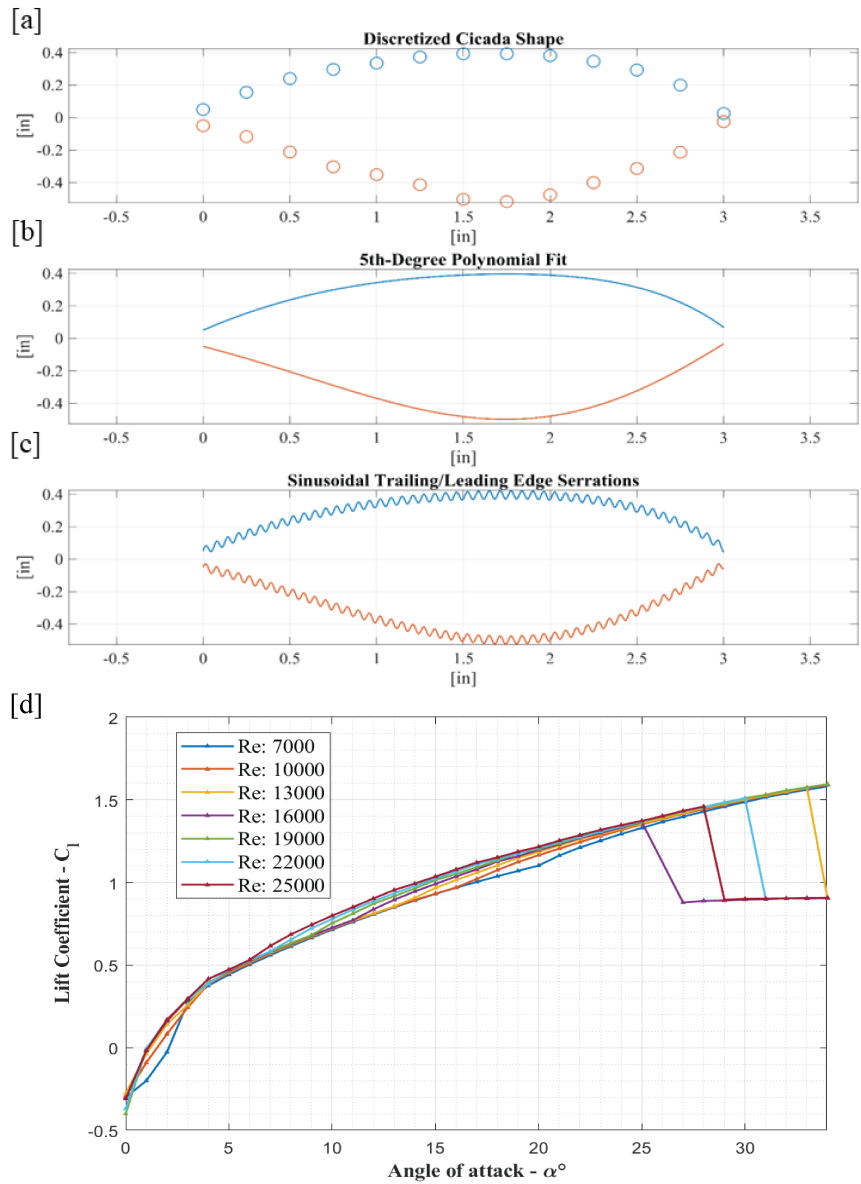

**Figure S1:** Cicada planform digitalization and 3D sinusoidal serration superposition. (a) Discrete points extracted from the outline of a cicada wing shape. (b) 5th-order fitted trailing-edge and leading-edge functions. (c) Superposition of the sinusoidal pattern. (d) NACA 8412 lift coefficient against the angle of attack curves under various Reynolds numbers.

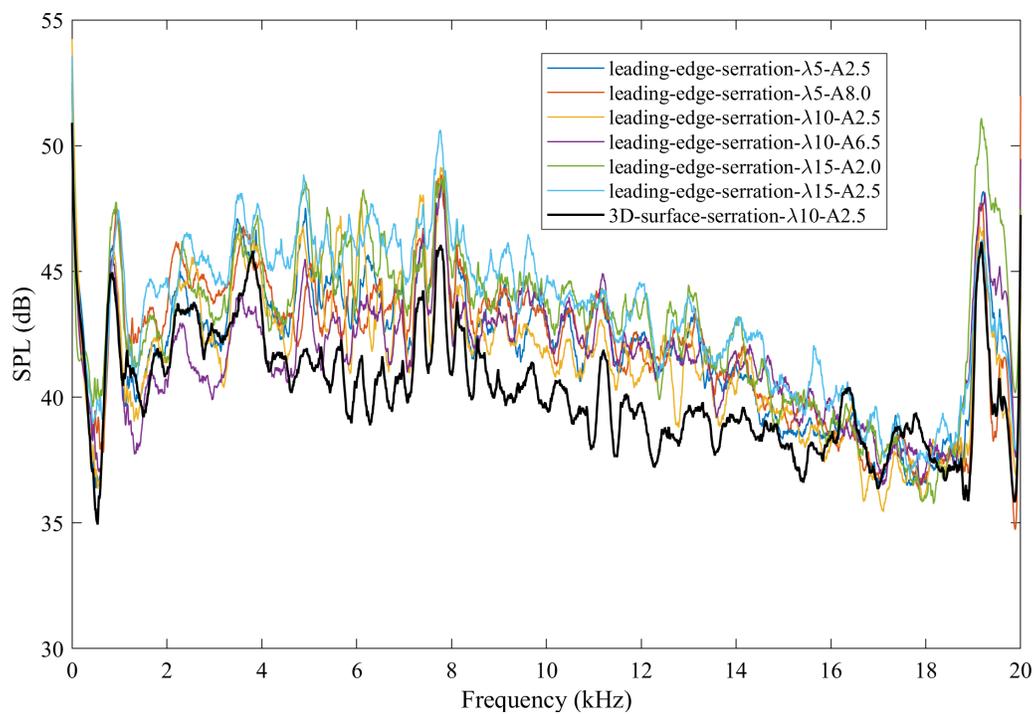

**Figure S2:** Sound spectrum plot of the 3D surface serrated and leading-edge serrated propellers.

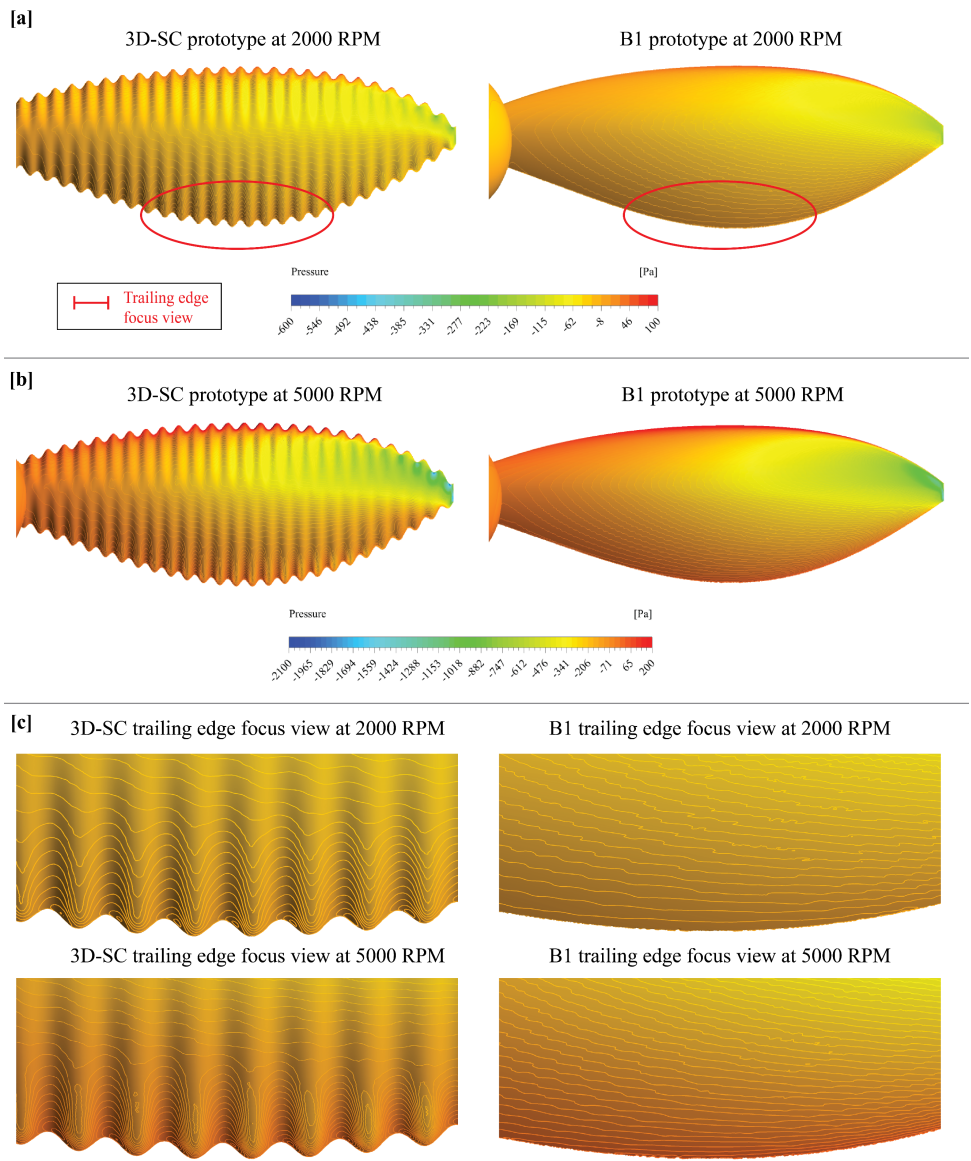

**Figure S3:** Gauge pressure contour. (a) surface gauge pressure contours of 3DSC (left) and B1 (right) prototypes at 2000 RPM. (b) surface gauge pressure contours at 2000 RPM. (c) focus view of the different pressure distribution patterns at 2000 and 5000 RPM.

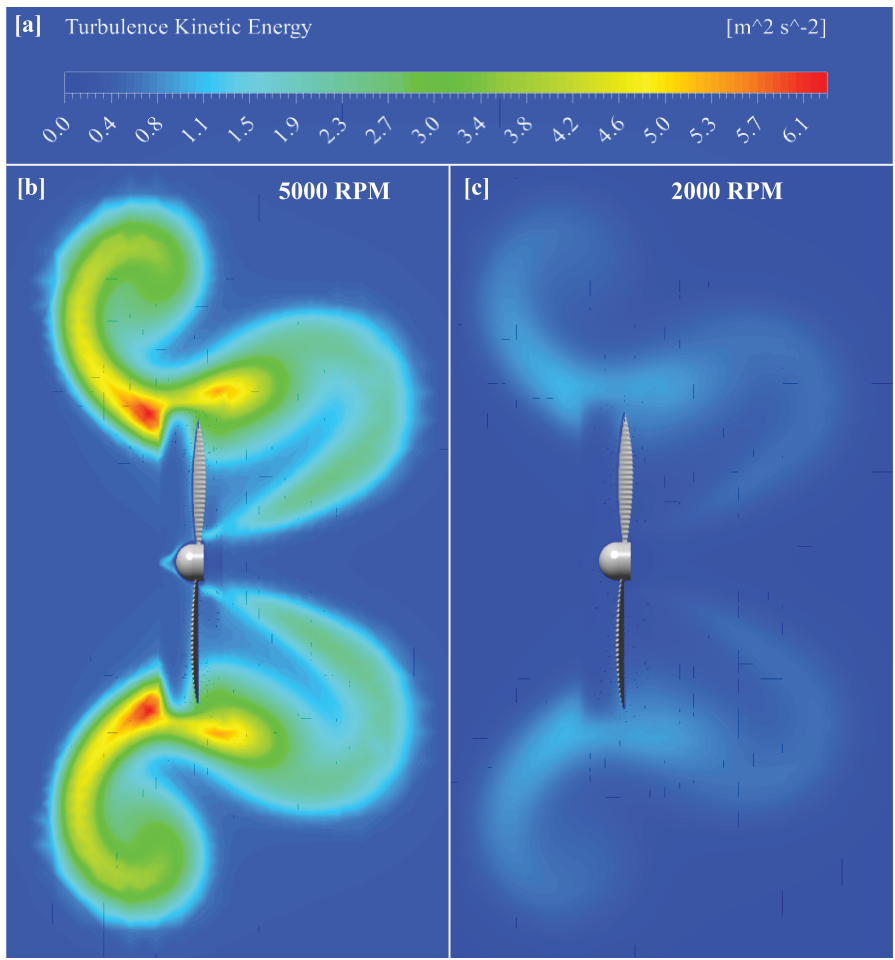

**Figure S4:** Turbulence kinetic energy contour. (a) contour legend, (b) 3D-SC propeller at 5000 RPM, (c) 3D-SC propeller at 2000 RPM

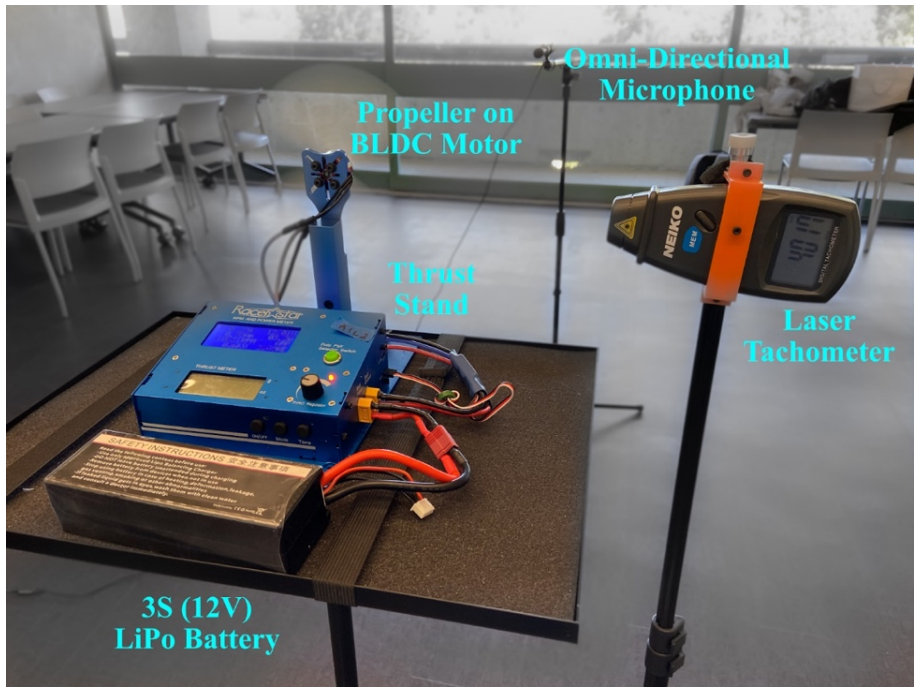

**Figure S5:** Experimental setup for sound and thrust data collection.

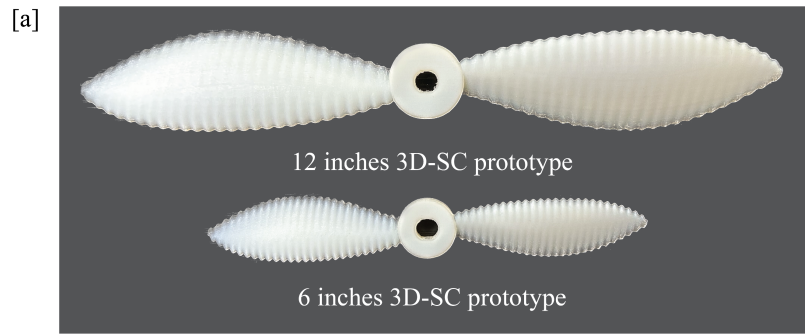

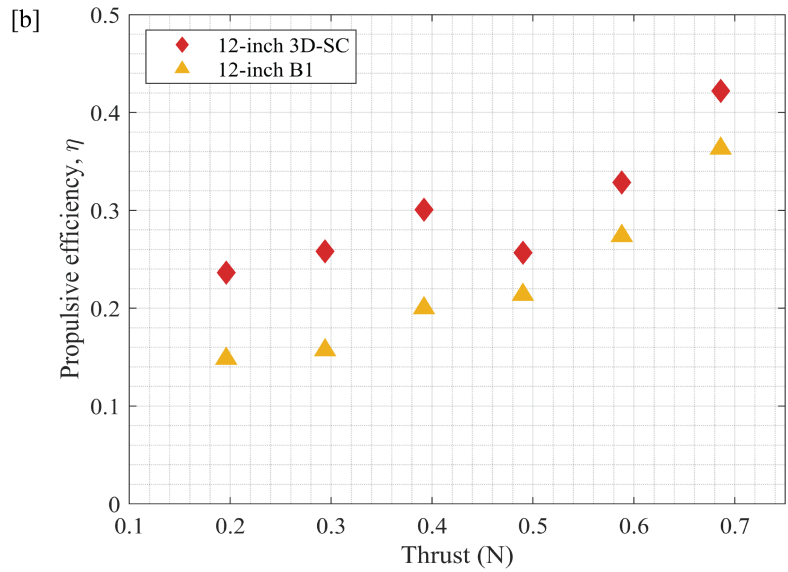

**Figure S6:** System efficiency of 12-inch 3D-SC and B1 prototypes. (a) comparison between 3D-SC propellers of different rotor diameters. (b) propulsive efficiency against thrust generation.

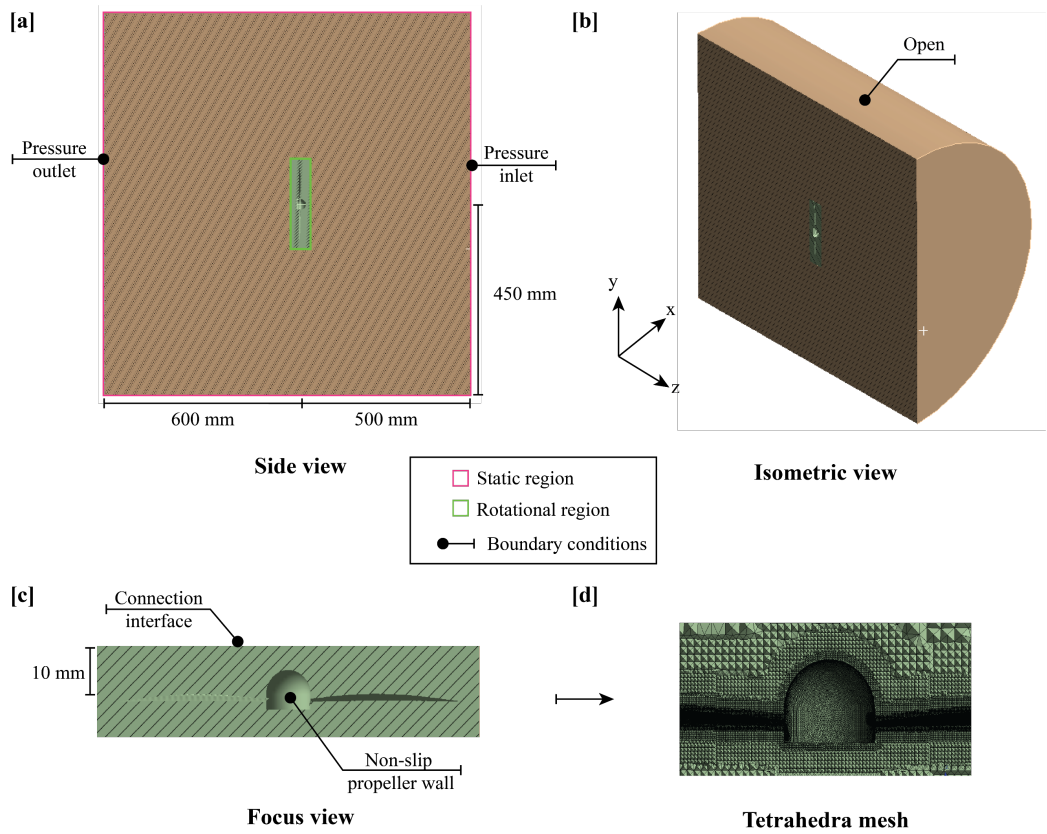

**Figure S7:** CFD simulation setup. (a) Side view of the fluid domain. (b) Isometric view of the fluid domain. (c) Focus view of the rotational region. (d) Depiction of the tetrahedra mesh setup.

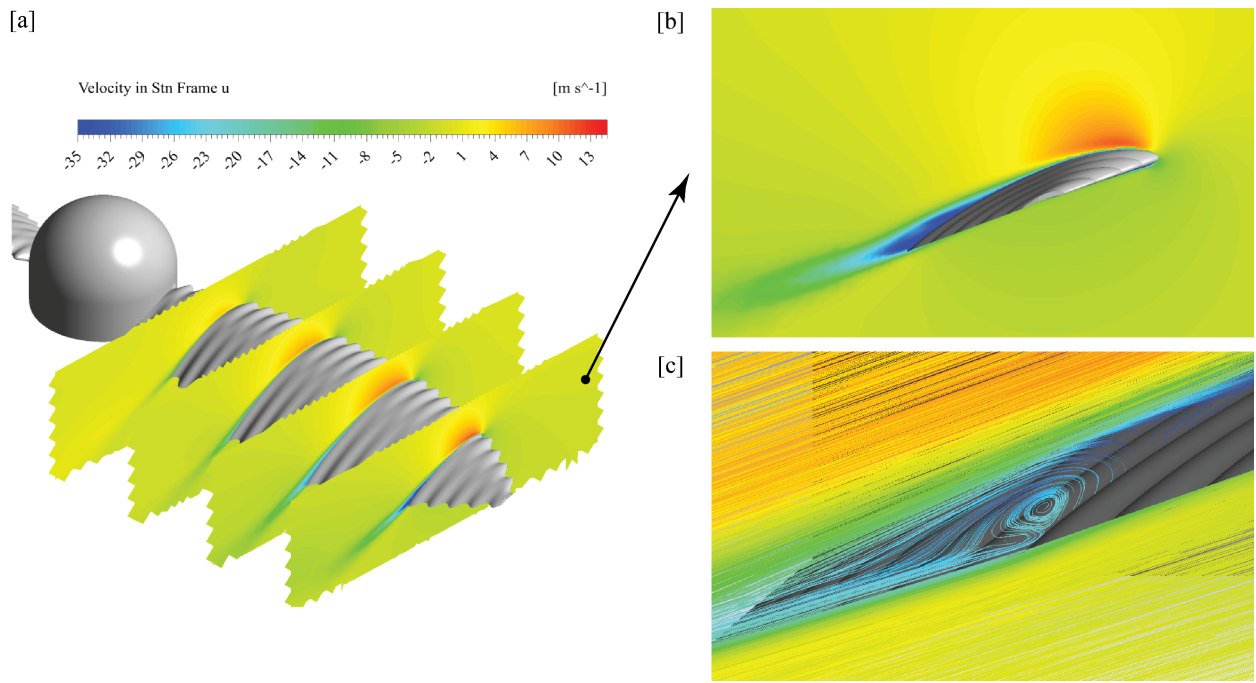

**Figure S8:** Income flow velocity contour. (a) velocity contour of chords located 0.2m, 0.35m, 0.5m, and 0.65m from the rotational center. (b) focus view of the velocity contour at 0.65m chord. (c) trailing edge flow separation visualization using the streamline plot.